\documentclass[12pt,a4paper]{article}
\usepackage{a4wide}
\usepackage{latexsym}
\usepackage{epsf}
\usepackage{amssymb}
\begin{document}
\def\be{\begin{equation}}
\def\ee{\end{equation}}
\def\bea{\begin{eqnarray}}
\def\eea{\end{eqnarray}}

\def\pd{\partial}
\def\a{\alpha}
\def\b{\beta}
\def\g{\gamma}
\def\d{\delta}
\def\m{\mu}
\def\n{\nu}
\def\t{\tau} 
\def\l{\lambda}
\def\s{\sigma}
\def\e{\epsilon}
\def\scri{\mathcal{J}}
\def\cM{\mathcal{M}}
\def\tcM{\tilde{\mathcal{M}}}
\def\RR{\mathbb{R}}
\def\CC{\mathbb{C}}

\hyphenation{re-pa-ra-me-tri-za-tion}
\hyphenation{trans-for-ma-tions}


\begin{flushright}
IFT-UAM/CSIC-00-29\\
hep-th/0010077\\
\end{flushright}

\vspace{1cm}

\begin{center}

{\bf\Large Sigma Model Corrections to the Confining Background}

\vspace{.5cm}

{\bf Enrique \'Alvarez}
\footnote{E-mail: {\tt enrique.alvarez@uam.es}},
{\bf Manuel Donaire}
\footnote{E-mail: {\tt mdx@adi.uam.es  }}
{\bf and Lorenzo Hern\'andez }
\footnote{E-mail: {\tt Lorenzo.Hernandez@uam.es }} \\
\vspace{.3cm}

\vskip 0.4cm

\ {\it Departamento de F\'{\i}sica Te\'orica, C-XI,
  Universidad Aut\'onoma de Madrid \\
  E-28049-Madrid, Spain \\and\\ 
Instituto de F\'{\i}sica Te\'orica, C-XVI,
  Universidad Aut\'onoma de Madrid \\
  E-28049-Madrid, Spain}\footnote{Unidad de Investigaci\'on Asociada
  al Centro de F\'{\i}sica Miguel Catal\'an (C.S.I.C.)}

\vskip 1cm


{\bf Abstract}

\end{center}

\begin{quote}
 Sigma model ($\a^{\prime}$) corrections to the confining string
background are obtained. The main result is that the Poincar\'e invariant 
ansatz
is maintained. Physical conditions for the dissapearance of the
naked singularity are discussed.
  
\end{quote}


\newpage

\setcounter{page}{1}
\setcounter{footnote}{1}
\section{Introduction}

In the paper \cite{alvarezgomez} a new confining string background has been proposed. This background
encodes in a precise way the renormalization group properties of the four dimensional gauge
theory, and it guarantees the vanishing of  the sigma model beta functions
 to $o(\alpha'\equiv l_s^{2})$.
\par
The explicit form of the background metric is:
\be\label{background}
ds^2 = g \, \eta_{\m\n}dx^{\m}dx^{\n} +l_c^2 dg^2 + \sum_{A=5}^{A=26}(dx^A)^2
\ee
where $l_c$ is a characteristic
length, unrelated {\em a priori} with the {\em string scale}, $l_s$. 
\par
This metric is a particular instance of the Poincar\'e invariant family
of metrics
\be\label{poinca}
ds^2 = a(\rho)\, \eta_{\m\n}dx^{\m}dx^{\n}+ d\rho^2 + \sum_{A,B=5}^{26} g_{AB}(x^A)dx^A dx^B
\ee
for which several interesting holographic properties (such as a C-theorem
and Callan-Symanzyk-like renormalization group equations) have been discussed
 in the literature \cite{poincare}.

The background dilaton reads
\be\label{}
\Phi = - log\,g
\ee

In terms of the usual renormalization group scale, $\mu$, the
dimensionless holographic variable is given by: 
\be\label{holographic}
g= \frac{g_0}{log\,\mu/\Lambda}
\ee
where $\Lambda$ is the renormalization group invariant mass scale of the gauge theory.
The ultraviolet (UV) region then corresponds to $g<<1$, whereas the infrared (IR) one
lies in $g>>1$;  by construction it does not make any sense to consider $\mu<\Lambda$,
which
would correspond to negative values of $g$ and imaginary dilaton fields.
\par
There is a question, however, about the range of the coordinates. We presumably want
 the $21$ {\em spectator} 
coordinates to be compactified on a very small torus, of common radius, say, $R$.

On the other hand, the background has got a singularity at $g=0$, but we obviously cannot trust
it that far, and for some purposes it is better to consider the (extensible) manifold obtained by
restriction to an interval $g\in (g_{UV},g_{IR})$.
\par

It is mathematically also possible to consider the solution extended to the real line, 
$g\in \mathbb{R}$, by symply putting absolute values:
\be\label{extension}
ds^2 = |g|\eta_{\m\n}dx^{\m}dx^{\n}  +l_c^2 dg^2 + \sum_{A=5}^{A=26}(dx^A)^2
\ee
and
\be\label{exten}
\Phi = - log\,|g|
\ee
in terms of the physical renormalization group scale, $\mu$, this means that we have extended
$\mu$ to the interval $(0,\Lambda)$ by using
\be\label{dual}
\mu\rightarrow\frac{\Lambda^2}{\mu}
\ee
\par
There are two physically different types of corrections to any physical quantity
evaluated in a given background: sigma model and stringy.
 Stringy corrections are 
proportional to $g_s$, which for us is exactly  the gauge coupling, $g$. Sigma model corrections
are proportional to $l_s^{2}$, and are the subject of the present paper.
\par
We expect roughly that sigma model corrections should become important when the curvature
is big, as measured in $l_s$ units, that is, for $R\sim \frac{1}{l_c^2 g^2}>>\frac{1}{l_s^2}$,
that is, for $g<<\frac{l_s^2}{l_c^2}$ (the UV region).
\par
 String corrections (to Green's functions or even to non local observables, such as Wilson loops),
on the other hand, are expected to become big when $g>>1$  (The IR region).
In fact both corrections are inextricably entangled through the soft dilaton theorem, as we have
emphasized in previous work.
\par

In the following, we are going to find that the solution above does indeed receive sigma model 
corrections;
of course this is enough to prove that it is not a coset model, which would instead be an exact solution
to the sigma model equations to all orders in $l_s^2/l_c^2 $. We have nevertheless included a 
direct proof in the appendix, putting the emphasis on coset models which in the semiclassical
approximation do enjoy the full Poincar\'e group as its isometry group.

\section{$l_s^4$ corrections}
So far we know that the string background  given by the formulae
\be\label{}
ds^2 = g  \eta_{\m\n}dx^{\m}dx^{\n}+l_c^2 dg^2 + \sum_{A=5}^{A=26}(dx^A)^2
\ee
\be
\Phi = - log\,g
\ee
satisfies the beta functions to O($\alpha'$).However we easily find that this
 solution is not exact. The Weyl anomaly coefficients to order
$o(l_s^4/l_c^4)$ read (see e.g. \cite{tseytlin}):

\bea
\beta^{\Phi}&&= \frac{D-26}{6}+l_s^2 \frac{(\nabla \Phi)^2 + \nabla^2 \Phi}{4} + l_s^4 R_{ABCD}R^{ABCD}= \frac{l_s^4}{l_c^4} \frac{5}{32g^4}\nonumber\\
\beta^{G}_{\m\n}&&=l_s^2  (R_{\m\n} - \nabla_{\m}\nabla_{\n}\Phi)
+ l_s^4 \frac{1}{2}R_{\m ABC} R_{\n}^{ABC}=\frac{l_s^4}{l_c^4} \frac{1}{4g^3}\eta_{\m\n}\nonumber\\
\beta^{G}_{44}&&=   l_s^2  (R_{44} - \nabla_{4}\nabla_{4}\Phi)
+ l_s^4 \frac{1}{2}R_{4 ABC} R_{4}^{ABC} = \frac{l_s^4}{l_c^2} \frac{1}{4g^4}
\eea
where  $A,B,C,\ldots=0 \ldots 25$, $\m,\n\ldots=0 \ldots 3$, and $x^4\equiv g$ denotes
the holographic coordinate.

At this point it is worth mentioning that all the results 
obtained are independent of the choice of the metric signature.

The simplest way to find a  background 
that enforces the vanishing of the beta functions to $o(l_s^4)$, 
starting with the unperturbed eq. (\ref{background}) and compatible with 
Poincar\'e invariance, consists of modifying the term  conformal
to the Minkowski metric in 
our background  by inserting an arbitrary function $f(g)$:
\be\label{a}
ds^2 = (g + \frac{l_s^2}{l_c^2} f(g)) \eta_{\m\n}dx^{\m}dx^{\n} +l_c^2 dg^2 + \sum_{A=5}^{A=26}(dx^A)^2
\ee
Note that  in this way  our dilaton remains unperturbed.
\par
The condition we get for the vanishing of the beta functions to $o(l_s^4)$ 
is the following:
\bea
\beta^{\phi }&&= [\frac{-f^{\prime}}{2g^2} + \frac{f}{2g^3} + 
\frac{5}{32g^4}]\frac{l_s^4}{l_c^4}  \nonumber\\
\beta^{G}_{\m\n}&&= [\frac{f}{2g^2} - 
\frac{f^{\prime}}{2g} - \frac{f^{\prime\prime}}{2} + 
\frac{1}{4l_c^2g^3}] \frac{l_s^4}{l_c^4}   \eta_{\m\n}\nonumber\\
\beta^{G}_{44}&&= [\frac{-2f}{g^3} + \frac{2f^{\prime}}{g^2} - \frac{2f^{\prime\prime}}{g} + \frac{1}{4l_c^2g^4}]\frac{l_s^4}{l_c^2} 
\eea

Unfortunately the system  above is algebraically incompatible .

We are then forced to make a 
more general perturbation on the metric, inserting two new  functions  
$f(g)$ and $w(g)$ in the expression: 
 
\be\label{perturbacion} 
ds^2= (g + \frac{l_s^2}{l_c^2} f(g)) \eta_{\m\n}dx^{\m}dx^{\n} + l_{c}^2(1 
 + \frac{l_s^2}{l_c^2} w(g))dg^2 +\sum_{A=5}^{A=26}(dx^A)^2
\ee 
but keeping (9) for the dilaton.
\par 
It is trivial that this is equivalent, up to a change  of
 coordinates, to   a background with $w(g)=1$, and a different $f'(g)$, as
well as  a new dilaton.
In terms of the function $h(g)$, such that $\frac{dh}{dg}= w(g)$, the new dilaton is given by:
\footnote{
The change of coordinates is given by
\be
g \longrightarrow g + \frac{l_s^2 h(g)}{2 l_c^2}
\ee
}
\be
\Phi = -log(g) +\frac{l_s^2}{l_c^2} D(g) 
\ee
where $D(g)=\frac{h(g)}{2 g}$, and the new function $f'(g)$ is given 
in terms of the old one by:
\be
 f'(g)\equiv f(g) - \frac{h(g)}{2}
\ee

\par

The new beta functions read:
\bea  
\beta^{\Phi} &&= \frac{l_s^4}{4l_c^4}(\frac{w^{\prime}}{2g} 
+ \frac{2f}{g^3} - \frac{2f^{\prime}}{g^2} + \frac{5}{8 g^4})\nonumber\\
\beta^{G}_{\m\n}&&=   
\frac{l_s^4}{l_c^4}(\frac{w^{\prime}}{4} + \frac{f}{2g^2} - 
\frac{f^{\prime}}{2g} - 
\frac{f^{\prime\prime}}{2} + \frac{1}{4 g^3})\eta_{\m\n}\nonumber\\
\beta^{G}_{44}&&= \frac{l_s^4}{l_c^2}(\frac{w^{\prime}}{2g} - 
\frac{2f}{g^3} + \frac{2f^{\prime}}{g^2} - 
\frac{2f^{\prime\prime}}{g} + \frac{1}{4 g^4}) 
\eea 
 
and they vanish when:
 
\bea 
f(g) &&= (c_{1}g - \frac{1}{32 g})\nonumber\\ 
w(g) &&=(c_{2} + \frac{1}{2g^2}) 
\eea 
 
Using now the relation (15) we may see this  solution in an equivalent way 
as a change on the dilaton field:   
\be 
D(g) = \frac{c_2}{2}  + \frac{c_3}{2g} - \frac{1}{4g^2} 
\ee 
Note that, in this manner, a new arbitrary constant $c_{3}$ arises. However, in this case, $c_2$ is nothing but the freedom we have to choose a zero point value on the original dilaton, due to the fact that it only appears  in the beta equations in the form of a derivative.\\

It is perhaps worth stressing that  we have verified  in passing 
that the only possible backgrounds $G_{\mu\nu}, B_{\mu\nu}$
preserving Poincar\'e invariance \footnote{Which in the case of
the Kalb-Ramond field (considered as a two-form) means that $\pounds(k) B= d C$.}, which, together with our dilaton 
$\Phi$ = - log\,$g$, saturate the   beta equations to order $o(l_s^2 /l_c^2)$ 
are of the form:
\be\label{teorema}
ds^2 = A_{1}g \eta_{\m\n}dx^{\m}dx^{\n} + A_{2}l_c^2 dg^2 + \sum_{A=5}^{A=26}(dx^A)^2
\ee
$A_{1}$,$A_{2}$  being arbitrary constants, wheras the Kalb-Ramond field
has to be trivial, $H=0$

It is easy to see that this family of solutions is equivalent to our initial metric up to a scale transformation on the coordinate $g$ as well as the addition of a proper constant to our original dilaton.

\section{The fate of the singularity}
Let us now consider the perturbed metric by itself:
\be\label{pert}
ds^2 = (g + (\frac{l_{s}}{l_c})^2(c_1g - \frac{1}{32 g})) 
\eta_{\m\n}dx^{\m}dx^{\n} + l_c^2 (1 + (\frac{l_{s}}{l_c})^2 (c_{2} + \frac{1}{2g^2}) ) dg^2 + \sum_{A=5}^{A=26}(dx^A)^2
\ee
We are now interested in searching for the possible  singularities of the metric that may 
have arisen after the $o(l_s^2/l_c^2)$ analytic perturbation.
\par
As usual, we find them by imposing the vanishing of the determinant of 
either the metric or its inverse.
From $ det\,g_{AB}$=0 we get:
\be\label{sing}
g_1^2 = l_{s}^2 \frac{1}{32(l_c^2 +l_{s}^2 c_1)}
\ee 

\be\label{singu}
g_2^2 = l_{s}^2 \frac{-1}{2(l_c^2 +l_{s}^2 c_2)}
\ee
 From  $det\,g^{AB}$=0 we obtain the original singularity $g_3^2$ = 0.
\par
How reliable are these putative singularities? First of all, in order for the whole 
perturbative expension to make sense, as we already said in the Introduction, 
the curvature has to be smaller than the string scale. 
This gives a condition for the  coordinate, namely,
\be
g \gg \frac{l_s}{l_c}
\ee
On the other hand, the curvature scalar of the new metric can be easily shown to
diverge when
\be\label{}
(-1+\frac{g^2}{g_1^2})(1-\frac{g^2}{g_2^2})=0
\ee
which clearly means that both $g=g_1$ and $g=g_2$ are true singularities; 
whereas $g=0$ would rather
become a Killing horizon.
\par
There are now two possibilities: If both $(1 + c_2 \frac{l_s^2}{l_c^2})>0$
and $(1 + c_1 \frac{l_s^2}{l_c^2})<0$ then $g_1$ as well as $ g_2$ are
 imaginary, so that the singularity has now been replaced by a horizon.
\par
It is worth remarking, however, that for this to be true, the constant
 $c_1$ has to be large, of order $o(1/\epsilon)$, where $\epsilon$
is the parameter of the perturbation, $\epsilon=\frac{l_s^2}{l_c^2}$,
which seems unnatural, although mathematically consistent.
\par
On the other hand, if $(1 + c_1 \frac{l_s^2}{l_c^2})<0$ there is a change of signature in the part of the
metric conformal to Minkowski space. 
If at least one of the above inequalities fails to be satisfied, then there is
a singularity at some positive value of $g$, and the would-be horizon 
remains hidden beyond the said singularity. The only effect of the 
perturbation has then been to shift the singularity a little bit on 
the real axis. 
\par
The new metric could be written in a form quite similar to the lowest 
order one, by defining the new variable \footnote{Determined from the condition
that $( 1 + (\frac{l_{s}}{l_c})^2 (c_{2} + \frac{1}{2g^2}) ) dg^2 
= d\tilde{g}^2$}
\be\label{cc}
\tilde{g}\equiv  g \sqrt{1 +\frac{l_s^2}{l_c^2} \,\left( 
c_2 + \frac{1}{2g^2} \right) } - 
     \frac{1}{\sqrt{2}}\sqrt{ \frac{l_s^2} {l_c^2}}\,    \log \left( \frac{1}{g\sqrt{2 \frac{l_s^2}{l_c^2}} } +  
          \frac{\sqrt{1 +\frac{l_s^2}{l_c^2} \left( c_2 + \frac{1}{2g^2} \right)} }{\frac{l_s^2}{l_c^2}}\right)
\ee
This yields the metric in the form
\be\label{poin}
ds^2 = f(\tilde{g})\eta_{\m\n}dx^{\m}dx^{\n}+ (d\tilde{g})^2 +\sum_{A=5}^{A=26}(dx^A)^2
\ee
with an adequate  $ f(\tilde{g})$. These coordinates are useful
only insofar as we restrict our interest to purely geometrical properties
of the background, because in them the dilaton gets also modified, so that
in general it is preferable to stick to the old system of coordinates
(in which the dilaton is {\em universal}).

\par
It is curious to observe that the Weyl tensor of the perturbed solution 
continues to vanish.
\section{Conclusions}
 We have calculated the corrections to the sigma model equations  (equivalent
to demanding vanishing Weyl anomaly coefficients) to order 
$o(l_s^4)$. 
\par
The first result is that, modulo convenient (and algebraically 
complicated) changes of coordinates (of the type of the one indicated 
in (\ref{cc}) above), the metric
can always be put in the general Poincar\'e invariant form (\ref{poinca})
, for 
which most convergence theorems have been proved, and which is conformally flat
(its Weyl tensor vanishes).
\par
This presumably means that many of the general results on the 
gravitational (holographic) interpretation of the renormalization 
group, such as the c-theorem, etc, \cite{poincare} when expressed in 
terms of the
conformal factor $a(\rho)$ will  remain valid, even after higher
terms in the sigma model expansion are considered.

\par
For generic values of the free parameters (integration constants) 
further results are not very spectacular; the position of the singularity is
simply shifted a tiny amount (proportional to the perturbation) on the
 real axis. But for
exceptional values of the parameters (namely, much bigger that the dimensionless strength of the perturbation) the singularity dissapears altogether 
and is replaced by a horizon. 
\par
Further computations are necessary before the 
physical reliability of the new horizon can be assessed.

\section*{Acknowledgments}
We have benefited from stimulating discussions with C\'esar G\'omez.
This work ~~has been partially supported by the
European Union TMR program FMRX-CT96-0012 {\sl Integrability,
  Non-perturbative Effects, and Symmetry in Quantum Field Theory} and
by the Spanish grant AEN96-1655.  The work of E.A.~has also been
supported by the European Union TMR program ERBFMRX-CT96-0090 {\sl 
Beyond the Standard model} 
 and  the Spanish grant  AEN96-1664.L.H. is supported by the spanish 
predoctoral grant AP99 43367460.


\section*{ Appendix Low dimension coset models with Poincar\' e isometry group}

It is well known that   GKO coset models \cite{goddard}
can be generally described by a gauged WZW model with group $G$,
 of which only the (vector action of the) subgroup $H\subset G$
has been gauged \cite{karabali}.
Besides, in the semiclassical regime (that is, when $k\rightarrow\infty$) , the gauge fields 
themselves (which
appear quadratically only) can be integrated,  yielding in this way an ordinary sigma model,
in a target space with dimension $d_{G/H}\equiv d_G - d_H$.
\par
In order for a given transformation to appear as a isometry in the semiclassical sigma model,
its action of the gauge fields must be trivial.
\par

If we want , for example, to obtain a  GKO coset, G/H with Poincar\'e invariance, then
in order for $ISO(3,1) \subset G$ to survive the vector gauging is necessary that
 $ ISO(3,1) \subset \cal Z$$[H]$ 
(the centralizer of $H$ in $G$). But this implies that $ H \subset \cal{Z}$$[ISO(3,1)]$ and, 
given the fact that  $ISO(3,1)$ has no center,  that $H$ has to be embedded in a subgroup 
generated by a basis $B \subset \mathfrak g$ out of $\mathfrak{iso}(3,1)$.

The scenario of lowest dimension (leaving aside the simplest
direct products of the form  $ISO(3,1)\otimes H / H$,
 with $H=U(1)$) consists of building G with Poincar\'e generators plus two extra ones, 
say $\{ Z,E \}$, $Z$ conmuting with all Poincar\'e generators but not with 
 $E$. Then we factor 
by $H=U(1)$ embedded in the $U(1)$ generated by $Z$.
\par
Independently of the different posibilities of selecting the conmutation relations 
of $E$ with $\mathfrak{iso}(3,1)$ algebra (compatible whith Jacobi identities), we expect
 the resulting coset to have dimension 11, provided that $H$ is not a null subgroup of $G$ (in 
this case an equivalence between vector and chiral gauge occurs accompained by an unexpected
 dimensional reduction of the GKO \cite{ardalan}).
\par 
In particular  this excludes the posibility of finding a five-dimensional GKO with Poincar\'e 
invariance (in fact, we can not find a (nontrivial) $d$ dimensional GKO with a symmetry group 
of dimension greater than $d-1$ provided the later has no center).
\par

If we consider for example the group as above generated by 
\be\label{} \nonumber
\{J_1,J_2,J_3,K_1,K_2, K_3,P_0,P_1,P_2,P_3,Z,E \}\nonumber
\ee \nonumber
 (with $J_i$ generators of rotations, 
$K_i$ boosts , $P_i$ translations \footnote{To be specific, the Poincar\'e
algebra in this basis is given by:
\bea
&&\left[J_i,J_j\right]=i \epsilon_{ijk}J_k\nonumber\\
&&\left[K_i,K_j\right]=-i\epsilon_{ijk}K_k\nonumber\\
&&\left[J_j,K_j\right]=i\epsilon_{ijk}K_k\nonumber\\
&&\left[J_i,P_j\right]=i\epsilon_{ijk}P_k\nonumber\\
&&\left[K_i,P_0\right]=-i P_i\nonumber\\ 
&&\left[K_i,P_j\right]=-i\delta_{ij}P_0
\eea
}
and the extra generators $Z$,$E$), then  there either $E$ itself 
(besides $Z$)  does not appear in the commutator $[E,Z]$ or it does .
In the first case, consistency with the Jacobi identities forces the 
subset $\{ Z,E \}$ 
to be a subalgebra:
\par

$[Z,E]=\alpha Z \ \ ,\ \  \alpha \neq 0$\\ \\
the conmutation relations of $E$ and $Z$ with $\mathfrak{iso}(3,1)$ given by:
\bea
&&\left[J_i,E\right]=A_{ij}J_j+B_{ij}P_j\nonumber\\
&&\left[K_i,E\right]=A_{ij}K_j+\frac{\epsilon_{ijk}}{2}B_{jk}P_0\nonumber\\
&&\left[P_i,E\right]=A_{ij}P_j\nonumber\\
&&\left[Z,E\right]=\a Z \nonumber
\eea
(where $A_{(ij)}=B_{(ij)}=0$).
But performing the redefinition:
\begin{displaymath}
E'=E -\frac{i}{2}\epsilon_{nml}B_{ml}P_n - \frac{i}{2}\epsilon_{nml}A_{ml}J_n
\end{displaymath}

it can be seen that this is nothing but the  direct sum of $\mathfrak{iso}(3,1)$ and the solvable algebra spanned by $\{Z,E'\}$.


In the second case, i.e., when $E$ belongs to the commutator of $[Z,E]$, it
can be shown that the subset $\{Z,E\}$ is not a subalgebra.
 The conmutation relations above keep the same values, but new structure 
constants show up: 
\begin{displaymath}
[E,Z]=\alpha Z + \beta E  -\frac{i}{2} \beta \epsilon_{nml} A_{ml} J_n  -\frac{i}{2} \beta \epsilon_{nml} B_{ml} Pn
\end{displaymath}
However, with the redefinition:
\begin{displaymath}
E' = E  - \frac{i}{2} \epsilon_{nml} A_{ml} J_n  -\frac{i}{2} \epsilon_{nml} B_{ml} P_n
\end{displaymath}
the full algebra can be written again as $ \mathfrak{iso}(3,1) \oplus \{Z,E'\}$as in the first case.
\par




\begin{thebibliography}{99}





\bibitem{poincare}
E.T.Akhmedov,
                  {\em A Remark on the AdS/CFT correspondence and the 
                  Renormalization Group Flow},
                  {\tt hep-th/9906217},Phys.Lett. B442 (1998) 152.\\
 E. \'Alvarez and C. G\'omez,
                {\em Geometric Holography, the 
                Renormalization Group and the c-Theorem},
                {\tt hep-th/9807226},Nucl.Phys. B541 (1999) 441-460.\\
A.~W.~Peet and J.~Polchinski,
{\em UV/IR relations in AdS dynamics,}
Phys.\ Rev.\  {\bf D59} (1999) 065011
{\tt hep-th/9809022}.\\
M.~Porrati and A.~Starinets,
{\em RG fixed points in supergravity duals of 4-d field theory and  asymptotically AdS spaces,}
Phys.\ Lett.\  {\bf B454} (1999) 77
[hep-th/9903085].\\
L.~Girardello, M.~Petrini, M.~Porrati and A.~Zaffaroni,
{\em Novel local CFT and exact results on perturbations of N = 4 super  
Yang-Mills from AdS dynamics}
JHEP {\bf 9812} (1998) 022
{\tt hep-th/9810126}.\\
V. Balasubramanian and P. Kraus,
                         {\em Spacetime and the Holographic 
                         Renormalization Group},
                         {\tt hep-th/9903190}.\\
D.~Z.~Freedman, S.~S.~Gubser, K.~Pilch and N.~P.~Warner,
{\em Renormalization group flows from holography supersymmetry and a  
c-theorem},
{\tt hep-th/9904017}\\
H.~Verlinde,
{\em Holography and compactification},
Nucl.\ Phys.\  {\bf B580} (2000) 264
{\em hep-th/9906182}.\\
J.~de Boer, E.~Verlinde and H.~Verlinde,
{\em On the holographic renormalization group},
JHEP {\bf 0008} (2000) 003
{\tt hep-th/9912012}.\\
E.~Verlinde and H.~Verlinde,
{\em RG-flow, gravity and the cosmological constant,}
JHEP {\bf 0005} (2000) 034
{\em hep-th/9912018}.\\






\bibitem{alvarezgomez} Enrique \'Alvarez and C\'esar G\'omez,{\em The Confining string
                                        from the Soft Dilaton Theorem},
                                       Nucl.Phys.B566:363-372,2000, 
                                      e-Print Archive: {\tt hep-th/9907158},\\
                                       {\em The Renormalization Group Approach to the Confining String} ,
                                        Nucl.Phys.B574:153-168,2000 ,
                                       e-Print Archive: {\tt hep-th/9911215}. 


\bibitem{ardalan} F. Ardalan,A.M. Ghezelbash,
                                      {\em Vector-Chiral Equivalence in Null Gauged WZNW Theory},
                                      Mod.Phys.Lett. A9 (1994)3749,
                                       e-Print Archive: {\tt  hep-th/9410158}. 


\bibitem{goddard} P.~Goddard, A.~Kent and D.~Olive,
                                       {\em Unitary Representations Of The Virasoro And Supervirasoro Algebras},
                                       Commun.\ Math.\ Phys.\  {\bf 103} (1986) 105.








\bibitem{karabali} D.~Karabali and H.~J.~Schnitzer,
                                      {\em Brst Quantization Of The Gauged WZW Action And Coset
                                      Conformal Field Theories},
                                       Nucl.\ Phys.\  {\bf B329} (1990) 649.



\bibitem{tseytlin} A. Tseytlin, {\em Sigma Model Approach to String Theory},
                                                             Int.J.Mod.Phys.A4:1257,1989 .



\end{thebibliography}
\end{document}